\documentclass[a4paper,10pt]{elsarticle}
\usepackage[utf8x]{inputenc}
\usepackage{xfrac}
\usepackage{amsmath}
\usepackage{color}
\usepackage{bm}
\usepackage{amssymb}
\usepackage{latexsym}
\usepackage{bbm}
\usepackage{ae}
\usepackage[T1]{fontenc}
\usepackage{upgreek}
\usepackage{graphicx}
\usepackage[english]{babel}
\usepackage{subfig}
\usepackage{xspace}
\usepackage{verbatim}
\usepackage{hyperref}

\newcommand{\micron}{\ensuremath{\upmu \text{m}}}

\newcommand{\mods}{\kappa_\mathrm{s}}
\newcommand{\modal}{\kappa_\alpha}
\newcommand{\modb}{\kappa_\mathrm{b}}
\newcommand{\Canum}{\ensuremath{Ca}\xspace}
\newcommand{\Ht}{\ensuremath{Ht}\xspace}

\newcommand{\ie}{\emph{i.e.}\xspace}
\newcommand{\eg}{\emph{e.g.}\xspace}

\begin{document}

 \title{Breakdown of deterministic lateral displacement efficiency for non-dilute suspensions: a numerical study}
 \author{R.~Vernekar}
 \author{T.~Krüger\corref{cor1}}
 \address{School of Engineering, The University of Edinburgh, Edinburgh EH9 3FB, United Kingdom}
 \cortext[cor1]{Corresponding author, email: timm.krueger@ed.ac.uk, phone: +44 131 650 5679, fax: +44 131 650 6554}

\begin{abstract}
 We investigate the effect of particle volume fraction on the efficiency of deterministic lateral displacement (DLD) devices.
 DLD is a popular passive sorting technique for microfluidic applications.
 Yet, it has been designed for treating dilute suspensions, and its efficiency for denser samples is not well known.
 We perform 3D simulations based on the immersed-boundary, lattice-Boltzmann and finite-element methods to model the flow of red blood cells (RBCs) in different DLD devices.
 We quantify the DLD efficiency in terms of appropriate ``failure'' probabilities and RBC counts in designated device outlets.
 Our main result is that the displacement mode breaks down upon an increase of RBC volume fraction, while the zigzag mode remains relatively robust.
 This suggests that the separation of larger particles (such as white blood cells) from a dense RBC background is simpler than separating smaller particles (such as platelets) from the same background.
 The observed breakdown stems from non-deterministic particle collisions interfering with the designed deterministic nature of DLD devices.
 Therefore, we postulate that dense suspension effects generally hamper efficient particle separation in devices based on deterministic principles.
\end{abstract}

  
\maketitle
  
\section{Introduction}
\label{section_introduction}

Separation of cellular blood components is an important step in clinical diagnosis of diseases, such as malaria~\cite{makler_review_1998}, as well as in medical research focusing on phenotype and/or genotype of the various subtypes of blood cells.
The traditional blood separation methods employed for clinical tests typically involve large sample volumes and often costly specialist equipment~\cite{makler_review_1998}.
The advent of microfluidic separation techniques for biological cells has opened up the possibility of replacement of traditional blood tests with lab-on-chip diagnostics~\cite{yager_microfluidic_2006, hou_microfluidic_2011}.
By scaling the separation process down to the cellular length scale one can reduce the sample volume and the time required for the tests. Furthermore, these microfluidic separation techniques lend easily toward downstream integration for analysis and diagnosis of cell populations of interest.

Deterministic lateral displacement (DLD) is a high-resolution and relatively straightforward size-based microfluidic separation technique first applied to the separation of hard polystyrene beads~\cite{huang_continuous_2004}.
It has the advantage of being label-free, relying solely on the device geometry without need for additional external forces to achieve separation (passive separation).
For this reason, it has been put to use in diverse separation applications, such as parasites from human blood~\cite{holm_separation_2011}, polystyrene micro-beads~\cite{loutherback_deterministic_2009}, purification of fungal spores~\cite{inglis_highly_2010}, separation of epithelial cells from fibroblast cells~\cite{green_deterministic_2009}, fractionation of human whole blood~\cite{davis_deterministic_2006}, removal of circulating tumor cells (CTCs) from blood~\cite{loutherback_deterministic_2012} and microfluidic droplet fractionation~\cite{joensson_droplet_2011}.
For a given geometry, a DLD device has two operation modes: ``displacement'', for larger particles, and ``zigzag'', for smaller particles as presented in section \ref{section_theory}.
Since DLD relies on \emph{deterministic} interactions between micro-obstacles and sample particles as they flow through the device, DLD devices are designed to operate under dilute conditions.
In this regime, the interaction of particles and obstacles is not much affected by particle-particle collisions so that particles can follow deterministic paths in the device.  

In recent years, the DLD technique has received much attention in the area of human blood cell separation.
In a previous work, the use of DLD has been demostrated for fractionating all cellular components of human blood~\cite{davis_deterministic_2006, mcgrath_deterministic_2014}.
With cell size as the sole criterion, DLD devices have been applied for separation of leukocytes from blood~\cite{davis_deterministic_2006, li_-chip_2007, zheng_deterministic_2005-1}, red blood cells (RBCs) from blood~\cite{davis_deterministic_2006, li_-chip_2007, zeming_rotational_2013}, platelets from blood~\cite{davis_deterministic_2006, inglis_microfluidic_2008} and even plasma from whole blood~\cite{davis_deterministic_2006}.
Furthermore, the use of DLD for deformability-dependent blood cell separation~\cite{kruger_deformability-based_2014, holmes_separation_2014} and fingerprinting~\cite{beech_sorting_2012, holmes_separation_2014} has been demonstrated.
In addition to fractionating blood constituents, external pathogens such as Trypanosomes have been separated from human blood using DLD~\cite{holm_separation_2011}.

Having achieved passive, label-free separation, those previous works show much promise for integration of the DLD principle in point-of-care blood diagnostic devices or even \emph{in vivo} use in human vasculature implants.
However, to the best of our knowledge, all the experiments use a diluted sample of blood.
The dilution is carried out either in the blood sample preparatory stage and/or by use of buffer streams in the device.
The blood cells are therefore always at a lower volume fraction than their natural haematocrit value.
Higher volume fractions close to the physiological haematocrit value of $\sim$40--45\% are desirable, though.
This would mean smaller sampling volumes, potentially faster results and minimum pre-treatment of the blood sample.
These factors become especially important when the species to be isolated (\eg~pathogens) are sparse and sample dilution would further reduce their already small concentration.

The performance of a DLD device at higher haematocrit values remains an open question and is a key requirement to get this technology into medical practice.
A few authors have only briefly looked at this matter and report that device clogging and separation breakdown due to cell deformability and stiction to device surfaces are some of the concerns~\cite{zheng_deterministic_2005-1, davis_deterministic_2006}.
To address this question, we require consideration from device manufacture as well as design aspects.
Going for an experimental evaluation of the DLD design is time-consuming and expensive due to the large number of different devices required.
At this point, simulations of cellular flow through DLD devices can reduce the workload since they allow for simple parameter variations.
Furthermore they provide unique insight into the deformation of cells, the resulting complex flow fields and the effect that these and other parameters (such as haematocrit) have on the device efficiency.

In this paper, we examine the effect of RBC volume fraction on the DLD performance using 3D computer simulations based on the immersed-boundary, lattice-Boltzmann and finite-element methods (section \ref{section_methods}).
RBCs are the most abundant cellular blood components (about 98\%) and therefore dominate the blood rheology.
Specifically, we analyse how the zigzag and displacement modes are affected by an increased number density of RBCs in the device.
We quantify the performance of these modes by defining appropriate failure probabilities.

One of our main results (section \ref{section_results}) is that the zigzag mode is relatively robust.
The displacement mode, however, is strongly affected by even moderate volume fractions (around 10--20\%).
We observe an eventual breakdown of the displacement mode at high volume fractions, rendering nearly all RBCs moving on zigzag-like trajectories.
This leads us to believe that it would probably be easier to separate a few larger cells from a dense RBC background than a few smaller particles from the same background.
Our results show that the failure probabilities have a different character in both device operation modes.
While failure events in the zigzag mode can annihilate each other, displacement failures always accumulate.
This explains why the zigzag mode is relatively robust upon haematocrit increase, while the displacement mode finally breaks down.

Our study provides valuable insight into the DLD behaviour at larger haematocrit, as summarised in section \ref{section_conclusions}.
The mechanism for the displacement breakdown is the increasing importance of \emph{non-deterministic} effects due to particle collisions.
Therefore, we expect those findings to apply to essentially all separation mechanisms relying on deterministic sorting of dense suspensions of particles of any kind.
The key question to make DLD devices more suitable for denser suspensions is how to reduce the failure probability in the displacement mode.
We believe that our study provides impetus to further research in the field.

\section{Methods and geometry}
\label{section_methods}

The employed numerical methods (section \ref{subsection_method}) and geometry (section \ref{subsection_geometry}) are the same as in \cite{kruger_deformability-based_2014}.
The major difference is the number of cells simulated simultaneously to vary the volume fraction (section \ref{subsection_simulations}).

\subsection{Numerical methods}
\label{subsection_method}

Several research groups have simulated RBC suspensions in the past years \cite{dupin_modeling_2007, macmeccan_simulating_2009, doddi_three-dimensional_2009, fedosov_predicting_2011, zhao_shear-induced_2011, freund_numerical_2014}.
Here, we use the lattice-Boltzmann method (LBM)~\cite{succi_lattice_2001, aidun_lattice-boltzmann_2010} for the fluid, the finite-element method (FEM) \cite{charrier_free_1989, shrivastava_large_1993} for the RBC membrane and the immersed-boundary method (IBM)~\cite{peskin_immersed_2002} for the fluid-membrane coupling.

We model the RBCs as biconcave closed membranes, meshed with 2,000 triangular surface elements.
The RBCs have a radius of $r = 3.9\, \micron$ along their major axis and a thickness of $2.4\, \micron$.
The elastic behaviour of the membrane is specified entirely by its shear, area dilation and bending moduli $\mods$, $\modal$ and $\modb$.
We keep the dimensionless value of $\modal = 0.5$ and $\modb / (\mods r^2) = 2.5 \cdot 10^{-3}$ constant, resulting in only one free parameter specifying the elastic behaviour of the RBCs.
The dimensionless interior and exterior fluid viscosities are $\eta = \frac{5}{6}$ and $\frac{1}{6}$, respectively.

All rigid boundaries (confining walls and obstacles of the DLD geometry) are modelled by the no-slip bounce-back method~\cite{ladd_numerical_1994}.
Membrane nodes experience a short-range repulsion near those boundaries and between each other to maintain a thin lubrication layer.
We employ periodic boundary conditions along the $y$-axis and shifted boundary conditions along the $x$-axis to realise a finite row shift.

A constant force density mimicking a pressure gradient drives the flow along the $x$-axis.
Additionally we impose a variable pressure gradient along the $y$-axis which counteracts any non-zero average flow along that axis.
The reason for this measure is that in real DLD devices there are also confining walls in $y$-direction (we assume they are far away but existing).
Due to the continuity equation and incompressibility condition, fluid cannot accumulate at those walls so that the average flow along the $y$-axis must vanish.

\subsection{Geometry}
\label{subsection_geometry}

\begin{figure}
 \centering
 \includegraphics[width=0.5\linewidth]{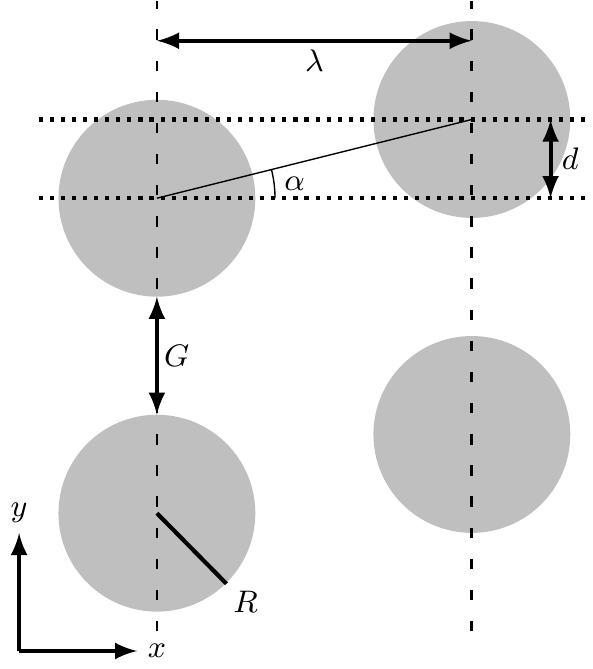} \\
 \caption{\label{fig:geometry2} Employed DLD geometry with pillar radius $R$, centre-to-centre distance $\uplambda$ and row shift $d$, which define the gap size $G$ and shift angle $\alpha$. Reprinted with permission from T.~Krüger, D.~Holmes, P.V.~Coveney, Biomicrofluidics, Vol. 8, Page 054114, (2014). Copyright 2014, American Institute of Physics.}
\end{figure}

We choose a shallow DLD device, with a height of $H = 4.8\, \micron$.
The large confinement along the $z$-axis forces the RBCs to align parallel with their major axis lying in the plane of the device ($x$-$y$-plane).
This arrangement ensures that the RBCs are not susceptible to orientation-dependent separation~\cite{beech_sorting_2012}.
The RBCs, however, remain deformable, which significantly influences their apparent size \cite{kruger_deformability-based_2014} and consequently their trajectories.

In accordance with most of the experimental work involving blood cells in DLD devices, we use a cylindrical pillar design for the obstacles as shown in Fig.~\ref{fig:geometry2}.
The pillar radius is $R = 10\, \micron$, and the centre-to-centre distance is $\uplambda = 32\, \micron$.
This leaves a gap size of $G = 12\, \micron$ in between adjacent pillars.
These device dimensions are kept invariant in all simulations.
The only free geometrical parameter is the row shift $d$, which defines the dimensionless displacement parameter $\epsilon = d / \uplambda$ and the angle $\alpha = \arctan \epsilon$.

We previously identified the streamline separation distances for the current DLD layout \cite{kruger_deformability-based_2014}.
The distances obtained for the row shifts $d$ used in the present work are collected in Table~\ref{tab:streamlines}.

\begin{table}
 \caption{\label{tab:streamlines} Streamline separation distances $s$ for selected row shift values $d$.}
 \centering
 \begin{tabular}{lll}
  \hline
  row shift & displacement parameter & separation distance \\
  $d\, [\micron]$ & $\epsilon$ & $s\, [\micron]$ \\
  \hline
  2.0 & $5/80$ & 1.2 \\
  4.4 & $11/80$ & 2.1 \\
  6.0 & $15/80$ & 2.7 \\
  \hline
 \end{tabular}
\end{table}

\subsection{Simulations}
\label{subsection_simulations}

We use the capillary number
\begin{equation}
 \Canum = \frac{p' \ell r}{\mods}
\end{equation}
as dimensionless measure for the RBC deformation in the ambient flow field.
The length $\ell = \sqrt{G H}$ is the geometric average of the width and height of the gap between two neighbouring pillars, and $p'$ is the pressure gradient along the $x$-axis.
Inertial effects are negligible \cite{kruger_deformability-based_2014}.

The simulation domain encloses a single pillar of the DLD device and consists of $80 \times 80 \times 12$ lattice units along the $x$-, $y$- and $z$-axes, respectively.
This gives a lattice constant of $\Delta x = 0.4\, \micron$ and an undeformed RBC diameter of $19.6\, \Delta x$.
By simulating only one pillar, we assume that the entire device is filled with RBCs at the specified volume fraction.
This simplification makes the simulations feasible in the first place, while capturing the essential DLD device properties \cite{kruger_deformability-based_2014}.

We previously mapped out the parameter space for deformability-based RBC separation \cite{kruger_deformability-based_2014}.
As a result, we obtained a line in the $\Canum$-$d$ parameter space which separates regions with RBCs moving on zigzag or displaced trajectories.
However, only a single RBC, \ie~the dilute limit, was considered.
Here, we use the known parameter map as a guide to investigate the effect of RBC volume fraction on the separation efficiency.

The relevant simulation parameters to explore are $d$ (device geometry), $\Canum$ (RBC deformability) and $\Ht$ (haematocrit, RBC volume fraction).
We consider five different $\Ht$-values ($8.0$, $16.1$, $32.2$, $40.2$ and $45.6\%$; the latter being close to the physiological value in humans).
Due to the periodic boundary conditions, we achieve these volume fractions with particle counts between 3 and 17.
All RBCs in a simulation have the same elastic properties.
The initial RBC positions and orientations are chosen arbitrarily, only taking care to avoid overlap with walls and other cells.
All simulations with identical \Ht share the same initial state.
For each considered volume fraction we have simulated three different row shifts ($d = 2.0$, $4.4$ and $6.0\, \micron$) and three different capillary numbers ($\Canum = 0.2$, $0.5$ and $1.0$) by varying the shear modulus $\mods$.
Therefore, we have run and analysed $5 \times 3 \times 3 = 45$ simulations in total.
The shear modulus for healthy RBCs is $\mods = 5.3\, \upmu\text{N} / \text{m}$.
Thus, in order to achieve \Canum between 0.2 and 1.0, we would require real-world pressure gradients between $35$ and $175\, \text{kPa}/\text{m}$.

\section{DLD operation modes}
\label{section_theory}

\begin{figure}
 \centering
 \includegraphics{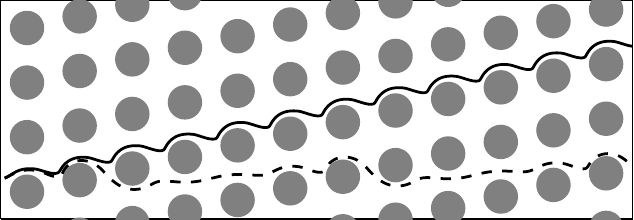}
 \caption{\label{fig_operationsmodes} Operation modes of a DLD device for particle separation from a background species (\eg~RBCs). In the displacement mode, the particles of the background species are forced on displaced trajectories (solid line) while the desired \emph{smaller} particles are expected to move horizontally on average. In the zigzag mode, the device is designed such that particles of the background species move on zigzag trajectories (dashed line) while the desired \emph{larger} particles move diagonally. The former mode may be used for separation of platelets from RBCs while the latter is usually employed for the separation of white blood cells from RBCs.}
\end{figure}

A DLD device designed to separate cells (\eg~white blood cells or platelets) from an RBC background can be classified into two cases (operation modes): one where the RBCs move in the displacement mode and the other where they are expected to move in the zigzag mode.
This is illustrated in Fig.~\ref{fig_operationsmodes}.
As we will demonstrate in the subsequent discussion, the behaviour of the cells in either mode is different from the other.
In view of this fact, we briefly discuss the displacement and zigzag modes separately.
In section \ref{section_results}, we will show how the RBC volume fraction affects both operation modes.

\begin{itemize}
 \item \textbf{Zigzag mode:}
 As the name suggests, in the zigzag mode a cell follows a zigzag path through the device as it moves downstream (dashed line in Fig.~\ref{fig_operationsmodes}).
 In this setup a single cell with a diameter smaller than the critical diameter travels along the flow streamlines, only moving around the pillars in order to compensate for the row shift.
 Thus, we expect the mean lateral displacement of the cell from its starting position to be zero at the outlet of a well designed DLD device.
 In order to achieve zero net displacement, a particle has to fall down to a lower lane (\ie~a diagonal along the pillars) every $n = \uplambda / d$ pillar encounters.
 This is nicely borne out in Fig.~\ref{fig_operationsmodes} (dashed line).
 For example, for a row shift $d = 6.0\, \micron$ and a centre-to-centre distance $\uplambda = 32\, \micron$, we find $n = 32 / 6.0 \approx 5.3$.
 \item \textbf{Displacement mode:}
 In the displacement mode the cell diameter is larger than the critical diameter.
 Therefore, the cell is laterally shifted at each pillar, along the array inclination, as it flows downstream in the device (solid line in Fig.~\ref{fig_operationsmodes}).
 Ideally, a particle in the displacement mode always remains in one single lane.
 In contrast to the zigzag mode, in this setup a cell is forced to interact with a pillar at every subsequent column.
 The expected lateral displacement is the product of the row shift $d$ and the number $N$ of pillars crossed.
\end{itemize}

\begin{figure}
 \centering
 \includegraphics{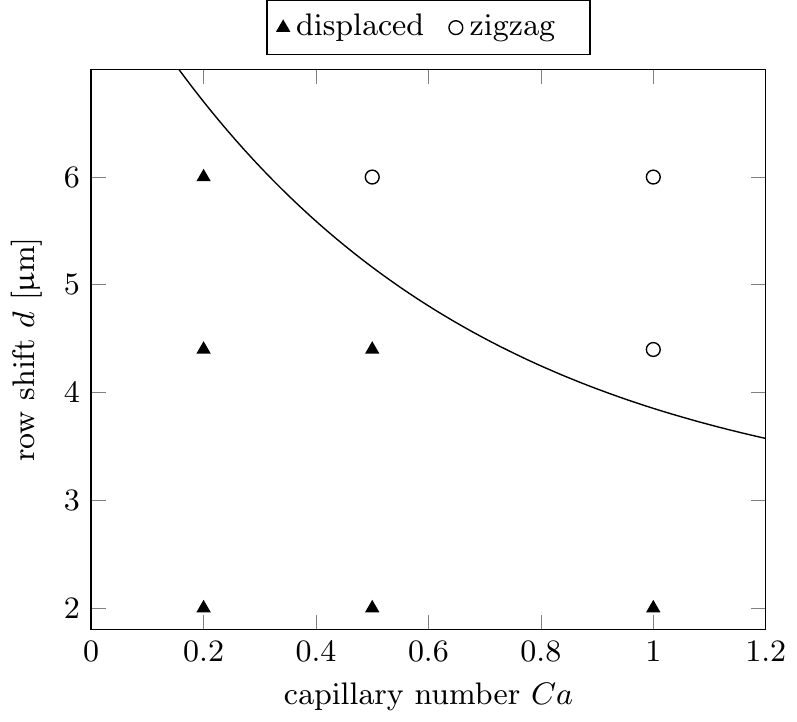}
 \caption{\label{fig_phasespace} Considered parameter space. We have simulated three different row shifts ($d = 2.0$, $4.4$, $6.0\, \micron$), each at three capillary numbers ($\Canum = 0.2$, $0.5$, $1.0$). Based on the results from \cite{kruger_deformability-based_2014}, we know that three of those data points lead to zigzag trajectories (circles) and the other six to displaced trajectories (triangles) in the dilute limit. The solid line indicates the separation of zigzag (above) and displaced trajectories (below), as identified in \cite{kruger_deformability-based_2014}.}
\end{figure}

Zigzag is the natural state in a DLD device, displacement is not.
On the one hand, fluid particles (\eg~water molecules) and other small particles always follow zigzag trajectories on average.
Their mean lateral motion is not affected by the presence of the obstacles, as long as the pressure gradient is along the device horizontal.
On the other hand, the displacement mode results from non-hydrodynamic volume exclusion effects forcing a sufficiently large particle to jump to another streamline at each obstacle.

Based on our previous analysis \cite{kruger_deformability-based_2014}, we know that the following parameters lead to zigzag trajectories of a single RBC: $(d, Ca) = (4.4\, \micron, 1.0)$, $(6.0\, \micron, 0.5)$ and $(6.0\, \micron, 1.0)$.
All other simulated parameter combinations lead to displaced trajectories.
These results are summarised in Fig.~\ref{fig_phasespace}.

\begin{figure}
 \centering
 \includegraphics[width=\linewidth]{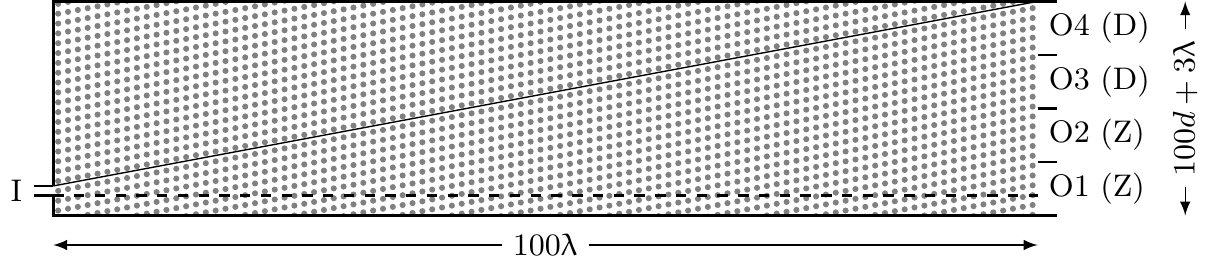}
 \caption{\label{fig_outlets} Definition of the DLD outlets. We define four imaginary equisized outlets at $x = 100 \uplambda$ (O1--O4). O1 and O2 are supposed to collect zigzag (Z) cells, O3 and O4 displaced (D) cells. The total outlet width is $100d + 3 \uplambda$ ($d / \uplambda = 0.1875$ for this figure). Note that we have extended the outlet width below the horizontal (dashed line) by $2\uplambda$ since a few cells tend to be negatively displaced (Fig.~\ref{fig:trajplot_vf45.6}). The solid diagonal line indicates the trajectory of an ideally displaced particle entering at the inlet (I) and staying in a single lane.}
\end{figure}

In order to analyse the performance of the DLD devices upon an increase in volume fraction, we define four imaginary outlets of equal width for collecting the RBCs after the 100\textsuperscript{th} encountered pillar ($N = 100$).
This arbitrary number is a compromise between the desired statistics and simulation runtime considerations.
The outlet layout in relation to the device geometry is shown in Fig.~\ref{fig_outlets}.
It is obvious that the total outlet width is different for different row shifts $d$ at the 100\textsuperscript{th} pillar.
We track individual RBC trajectories and make sure that the simulations run until all cells have encountered at least 100 pillars ($3.2\, \text{mm}$ downstream distance).

We emphasise that the above outlet definition is convenient for our analysis, but it does not necessarily reflect the requirements of a typical experiment.
Real devices may have fewer or more outlets or outlets with different sizes or outlets after a different number of pillars.
This, however, does not make our observations and conclusions any less relevant.

\section{Results and discussion}
\label{section_results}

We present our numerical results with a particular focus on the effect of the volume fraction $Ht$.
After discussing the RBC trajectories in section \ref{section_trajectories}, we introduce and analyse the failure event probabilities of RBCs (section \ref{section_failure}).
Finally, we investigate the spatial distribution of RBCs in the outlets of the DLD device (section \ref{section_outlets}).

\subsection{Cell trajectories}
\label{section_trajectories}

\begin{figure}
 \subfloat[$d = 2.0\,\micron$, $Ca = 0.2$]{\includegraphics[width=0.475\linewidth]{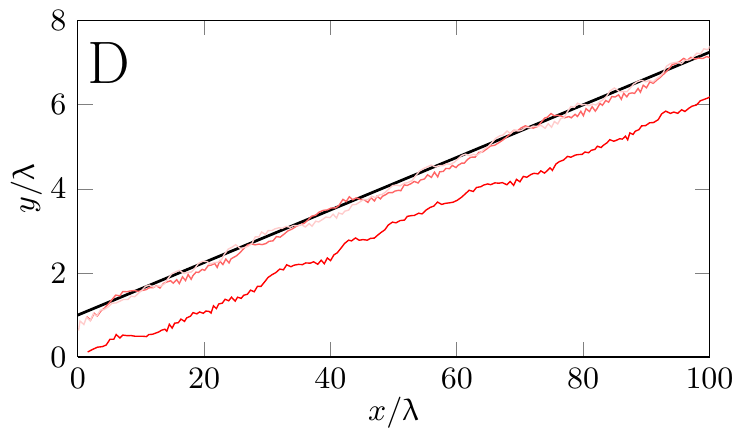}} \hfill
 \subfloat[$d = 2.0\,\micron$, $Ca = 1.0$]{\includegraphics[width=0.475\linewidth]{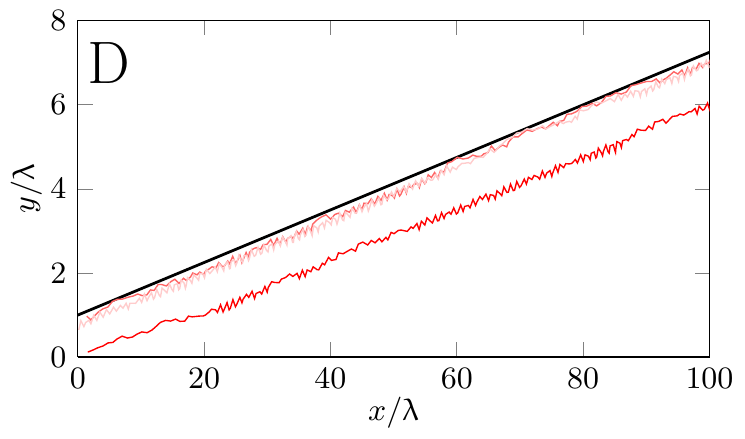}} \\
 \subfloat[$d = 4.4\,\micron$, $Ca = 0.2$]{\includegraphics[width=0.475\linewidth]{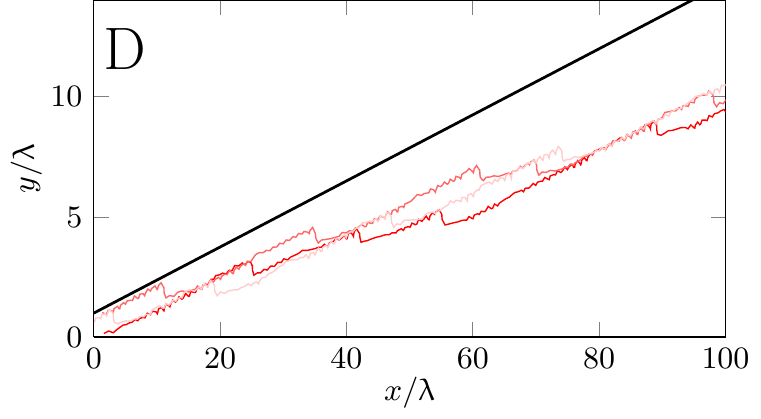}} \hfill
 \subfloat[$d = 4.4\,\micron$, $Ca = 1.0$]{\includegraphics[width=0.475\linewidth]{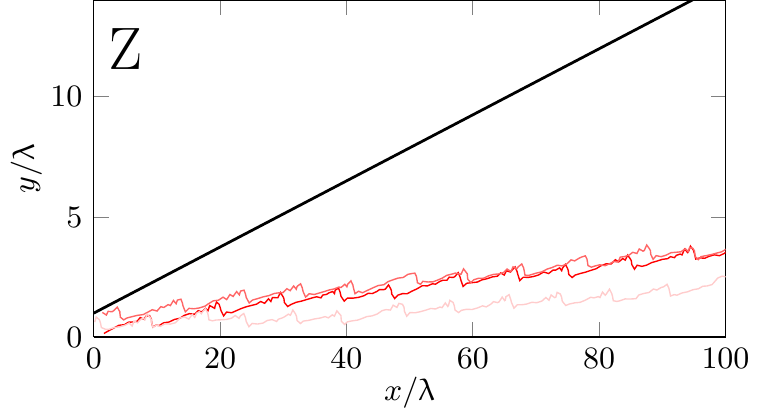}} \\
 \subfloat[$d = 6.0\,\micron$, $Ca = 0.2$]{\includegraphics[width=0.475\linewidth]{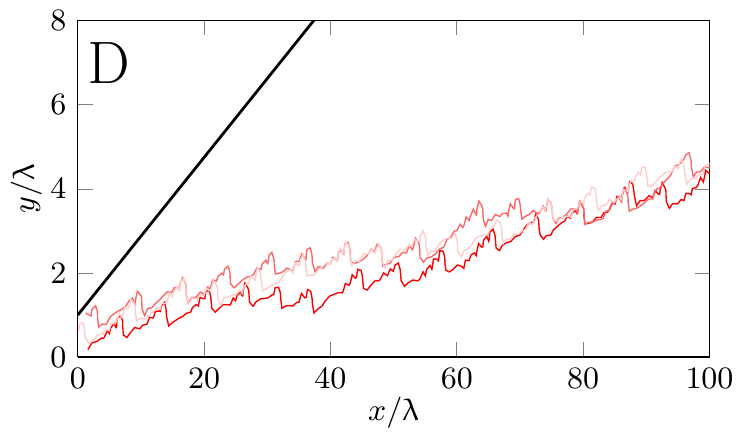}} \hfill
 \subfloat[$d = 6.0\,\micron$, $Ca = 1.0$]{\includegraphics[width=0.475\linewidth]{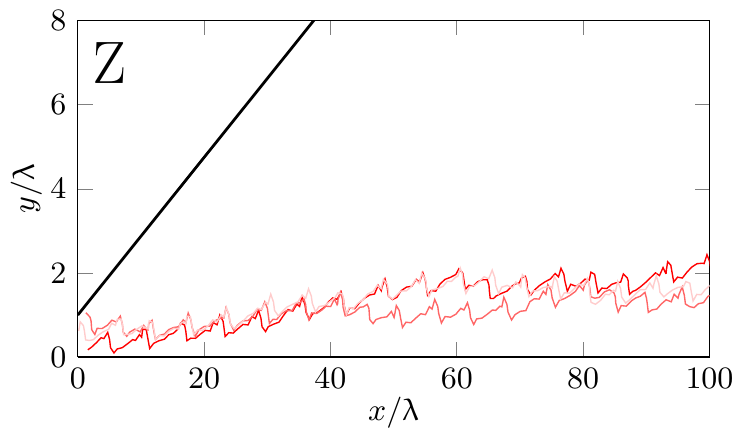}} \\
 \caption{\label{fig:trajplot_vf8.0} Cell trajectories (different red shades for individual cells) for $Ht = 8.0\%$. The black solid line indicates the inclination of the DLD lanes. Expected operation modes are designated by D for displacement or Z for zigzag according to Fig.~\ref{fig_phasespace}. Both axes are normalised by the pillar-to-pillar distance $\uplambda$. Note that the $x$- and $y$-axes are not shown to scale.}
\end{figure}

\begin{figure}
 \subfloat[$d = 2.0\,\micron$, $Ca = 0.2$]{\includegraphics[width=0.475\linewidth]{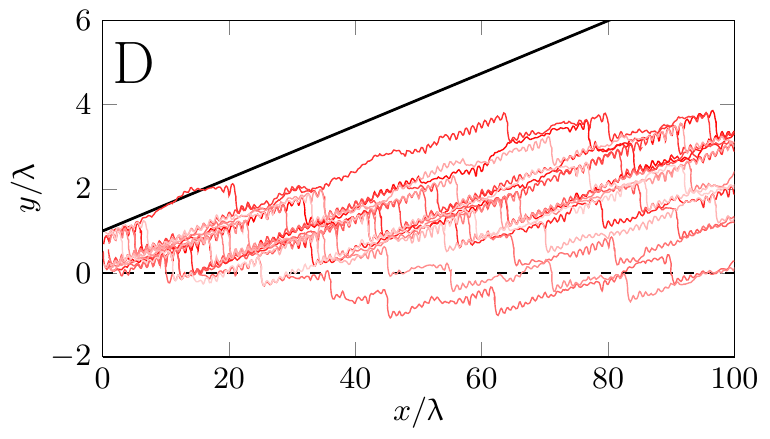}} \hfill
 \subfloat[$d = 2.0\,\micron$, $Ca = 1.0$]{\includegraphics[width=0.475\linewidth]{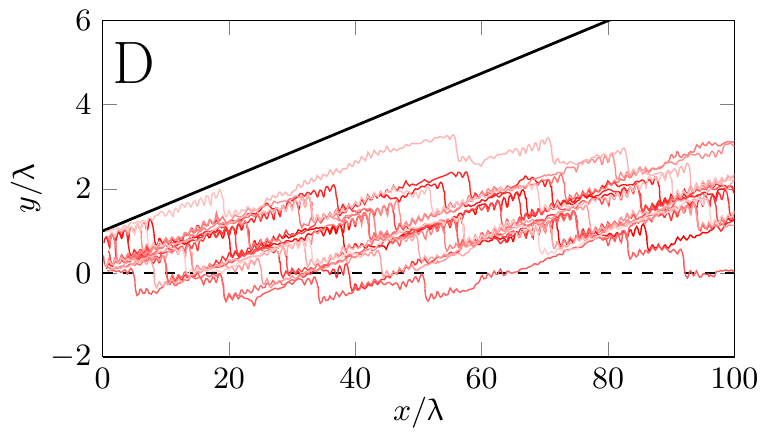}} \\
 \subfloat[$d = 4.4\,\micron$, $Ca = 0.2$]{\includegraphics[width=0.475\linewidth]{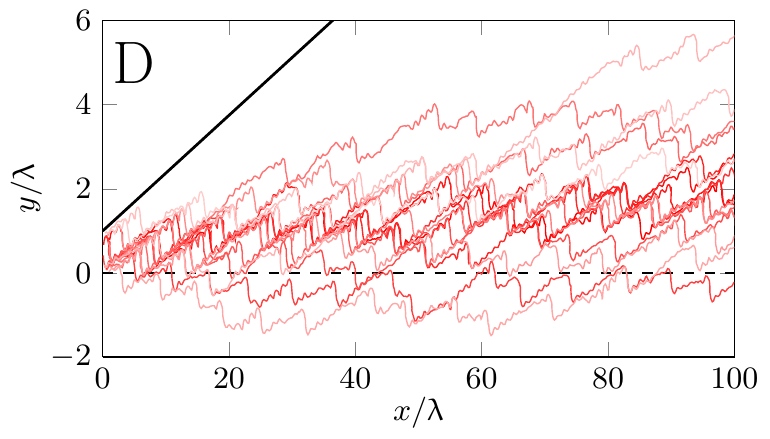}} \hfill
 \subfloat[$d = 4.4\,\micron$, $Ca = 1.0$]{\includegraphics[width=0.475\linewidth]{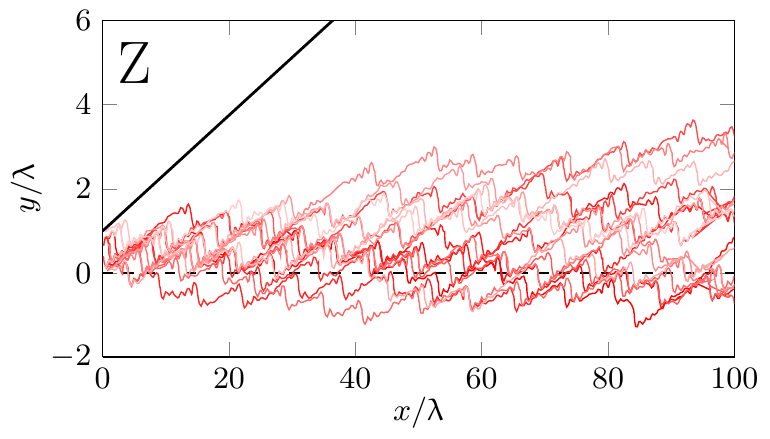}} \\
 \subfloat[$d = 6.0\,\micron$, $Ca = 0.2$]{\includegraphics[width=0.475\linewidth]{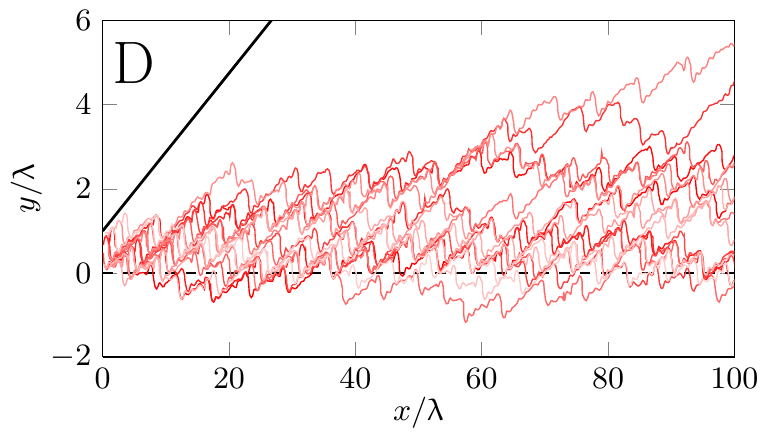}} \hfill
 \subfloat[$d = 6.0\,\micron$, $Ca = 1.0$]{\includegraphics[width=0.475\linewidth]{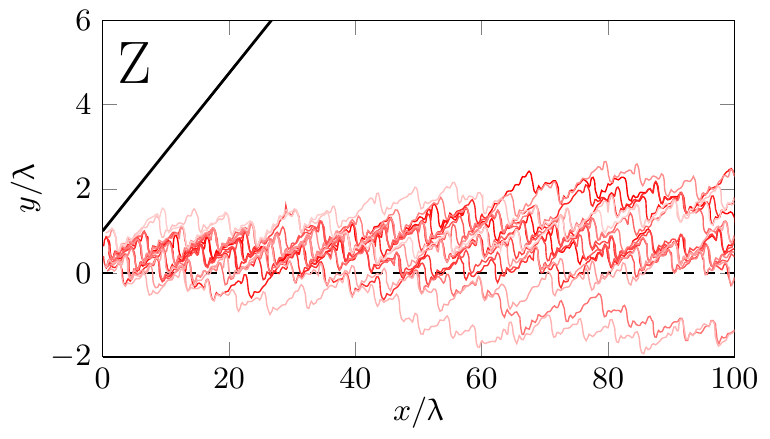}} \\
 \caption{\label{fig:trajplot_vf45.6} Cell trajectories (different red shades for individual cells) for $Ht = 45.6\%$. The dashed line indicates the device horizontal. See Fig.~\ref{fig:trajplot_vf8.0} for further explanations. Note that the $x$- and $y$-axes are not shown to scale.}
\end{figure}

Fig.~\ref{fig:trajplot_vf8.0} and Fig.~\ref{fig:trajplot_vf45.6} depict the RBC trajectories in the DLD device.
We show representative data for two different volume fractions: $Ht = 8.0\%$ (3 simulated cells) in Fig.~\ref{fig:trajplot_vf8.0} and $Ht = 45.6\%$ (17 simulated cells), which is close to the physiological value, in Fig.~\ref{fig:trajplot_vf45.6}.
The $x$- and $y$-axes are normalised by the pillar-to-pillar distance $\uplambda$ so that the label on the $x$-axis indicates the number of pillars passed by the cells.
Both figures consist of a set of subfigures with different row shifts ($d = 2.0$, $4.4$, $6.0\,\micron$, varied row-wise) and different capillary numbers ($Ca = 0.2$, $1.0$, varied column-wise).
The expected operational mode (Z for zigzag or D for displacement, \emph{cf.}~Fig.~\ref{fig_phasespace}) for a single RBC is marked in each subfigure.
Small jumps along the $y$-axis indicate cells passing over and above a pillar.
Sharp intermittent dips in negative $y$-direction show how cells pass below a pillar and fall down to another lane.

From Fig.~\ref{fig:trajplot_vf8.0} and Fig.~\ref{fig:trajplot_vf45.6} we see that, except for $d = 2.0\,\micron$ at $Ht = 8.0\%$, all RBCs fail to follow their designated trajectories: cells are neither in a pure displacement nor a pure zigzag mode.
At $Ht = 45.6\%$, the cells display a large scatter about their mean lateral position.
Since cell scatter, which is caused by cell collisions, is not deterministic, it is an indicator for a reduced DLD predictability and efficiency.
At the outlet ($x / \uplambda = 100$), the maximum cell scatter is as large as $\approx 7 \uplambda$ for $Ht = 45.6\%$ (Fig.~\ref{fig:trajplot_vf45.6}(c,e)), whereas for $Ht = 8.0\%$ the cell scatter is $\approx \uplambda$ and therefore relatively small (Fig.~\ref{fig:trajplot_vf8.0}).

An increase in volume fraction results in cells deviating from their expected ideal mean trajectory that is followed by a single cell in the device under the same conditions.
A DLD operation mode can be termed ``robust'' if most of the cells end up in their designated outlets and cellular separation can therefore be achieved.
We obseve that, in the zigzag mode, the mean outlet cell position does not deviate more than $\approx 3 \uplambda$ from the initial position.
Therefore, the zigzag mode is relatively robust upon an increase of volume fraction.
However, the lateral mean position in displacement mode is far from the expected one and varies with $d$ and $Ca$.
This means that the displacement mode is not robust.

Based on the phase space in Fig.~\ref{fig_phasespace}, we expect the zigzag mode for $Ht = 45.6\%$ with $d = 4.4$, $6.0\,\micron$ and $Ca = 1.0$ (Fig.~\ref{fig:trajplot_vf45.6}(d,f)).
Indeed we observe only relatively small undesired lateral cell displacements.
On average, these cells travel along near horizontal trajectories and could be collected within the first device outlet.
This leaves device outlets farther away from the horizontal free for collection of larger sized particles (\eg~white blood cells).

The situation is different for the expected displacement mode.
In Fig.~\ref{fig:trajplot_vf45.6}(a--c,e) we observe that the displacement mode deteriorates significantly with increased volume fraction.
In fact, at $Ht = 45.6\%$, the mean cell trajectory direction is close to the horizontal.
This would make it difficult to separate smaller particles (\eg~platelets) from a background of dense RBCs.

\subsection{Failure rates}
\label{section_failure}

We define a \emph{failure} as an event when a cell takes a ``wrong turn''.
For example, a failure in the displacement mode (displacement failure) means that a cell is not bumped up and changes the streamline but drops to a lower lane.
Each failure is associated with a lateral displacement penalty.
A displacement failure leads to a particle being located by $\uplambda$ lower than its expected lateral position, so $-\uplambda$ is the lateral penalty for a single displacement failure.

The zigzag mode behaves differently because a particle on a zigzag trajectory experiences a number of displacement events followed by one zigzag event.
Therefore, there are two different failure types in the zigzag mode: one where a cell is bumbed up once too often, the other when it falls to a lower lane once too often.
In any case, each failure leads to an unexpected lateral cell displacement by about $\pm \uplambda$, so the penalty is the same, up to its sign, for all failure modes.

The crucial point is that both zigzag failure modes tend to cancel each other because they lead to an unexpected displacement in \emph{different directions}.
One occurrence of the first failure mode compensates for one occurrence of the other failure mode.
This is completely different for the displacement mode where cells can only drop to a lower lane, but we have never observed a cell jumping an entire lane up.
Therefore, displacement failures always accumulate while zigzag failures can annihilate each other.
This makes the displacement mode susceptible to a breakdown when the failure rate increases.
Obviously, as shown in section \ref{section_trajectories}, this happens for increasing volume fractions where cell collisions lead to less deterministic behaviour.

Our observation suggests that the zigzag mode survives at higher volume fractions because both failure modes approximately cancel each other on average.
However, individual cells still behave differently.
Some experience more failures with a $+\uplambda$ penalty, others more with a $-\uplambda$ penalty.
This leads to the lateral cell scatter in the trajectory plots (Fig.~\ref{fig:trajplot_vf45.6}(d,f)).

\subsubsection{Zigzag failure}
\label{section_zigzag}

In order to quantify the effect of the device parameters on the cell trajectories, we define failure probabilities.
Let $\Delta_y$ be the lateral displacement error for a cell in the zigzag mode at the $N^\text{th}$ column.
This means that the cell is laterally displaced by $\Delta_y$ compared to its expected position.
A cell travelling on its expected trajectory obviously yields $\Delta_y = 0$.
For cells moving too far up (positive $y$-direction) we get $\Delta_y > 0$, for those moving too far down we have $\Delta_y < 0$.

Introducing failure probabilities $p^+_\text{f,Z}$ and $p^-_\text{f,Z}$ for both zigzag failure modes ($p^+_\text{f,Z}$ corresponds to an event where the cell is bumped above a pillar once too often, which leads to a positive lateral shift, while $p^-_\text{f,Z}$ indicates that the cell falls down to a lower lane once too often), we can write
\begin{equation}
 \label{equation_zigzag_failure}
 \Delta_y = N \uplambda \left(p^+_\text{f,Z} - p^-_\text{f,Z}\right).
\end{equation}
This equation means that a cell may experience a failure at each of the $N$ columns with probability $p^\pm_\text{f,Z}$.
Each failure contributes with $\pm \uplambda$ to the displacement $\Delta_y$ as discussed in section \ref{section_trajectories}.
Note that both zigzag failure modes tend to cancel each other, so only the net failure probability $p_\text{f,Z} = p^+_\text{f,Z} - p^-_\text{f,Z}$ will affect the final lateral cell position.
The probability $p_\text{f,Z}$ is a measure of the intrinsic failure mechanism for a given device under the given flow conditions ($Ca$, $Ht$).

In the zigzag mode, let $N_\text{up}$ and $N_\text{dn}$ be the number of times a cell chooses to move above or below a pillar obstacle encountered in the flow.
Furthermore, let $N_\text{up}^0$ and $N_\text{dn}^0$ be the number of times a single cell would flow above or below a pillar obstacle in an ideal zigzag mode.
It is clear that the total number of events equals the number $N$ of pillars encountered in any case:
\begin{equation}
 N = N_\text{dn}^0 + N_\text{up}^0 = N_\text{dn} + N_\text{up}.
\end{equation}
Furthermore, we know from the zigzag design requirements that
\begin{equation}
 \frac{N_\text{dn}^0}{N} = \frac{d}{\uplambda} = \frac{1}{n},
\end{equation}
where $n$ is the period for the row shift of the device.
The last equation reflects the fact that on average a cell on a perfect zigzag trajectory does not have a lateral net displacement.
In other words, the ratio of $N_\text{dn}^0$ to $N_\text{up}^0$ is uniquely determined by the device geometry ($d$ and $\uplambda$).

In practice, we calculate the net zigzag failure probability $p_\text{f,Z}$ for each individual cell \emph{via} eq.~\eqref{equation_zigzag_failure} and the observed values for $\Delta_y$ at $x = 100 \uplambda$ ($N = 100$).
This allows us to compute the average and variance over all the cells in a particular simulation.
In terms of number of events, we can also write
\begin{equation}
 p_\text{f,Z} = \frac{N_\text{dn}^0 - N_\text{dn}}{N} = \frac{N_\text{up} - N_\text{up}^0}{N}.
\end{equation}

\begin{figure}
 \centering
 \includegraphics[width=0.8\linewidth]{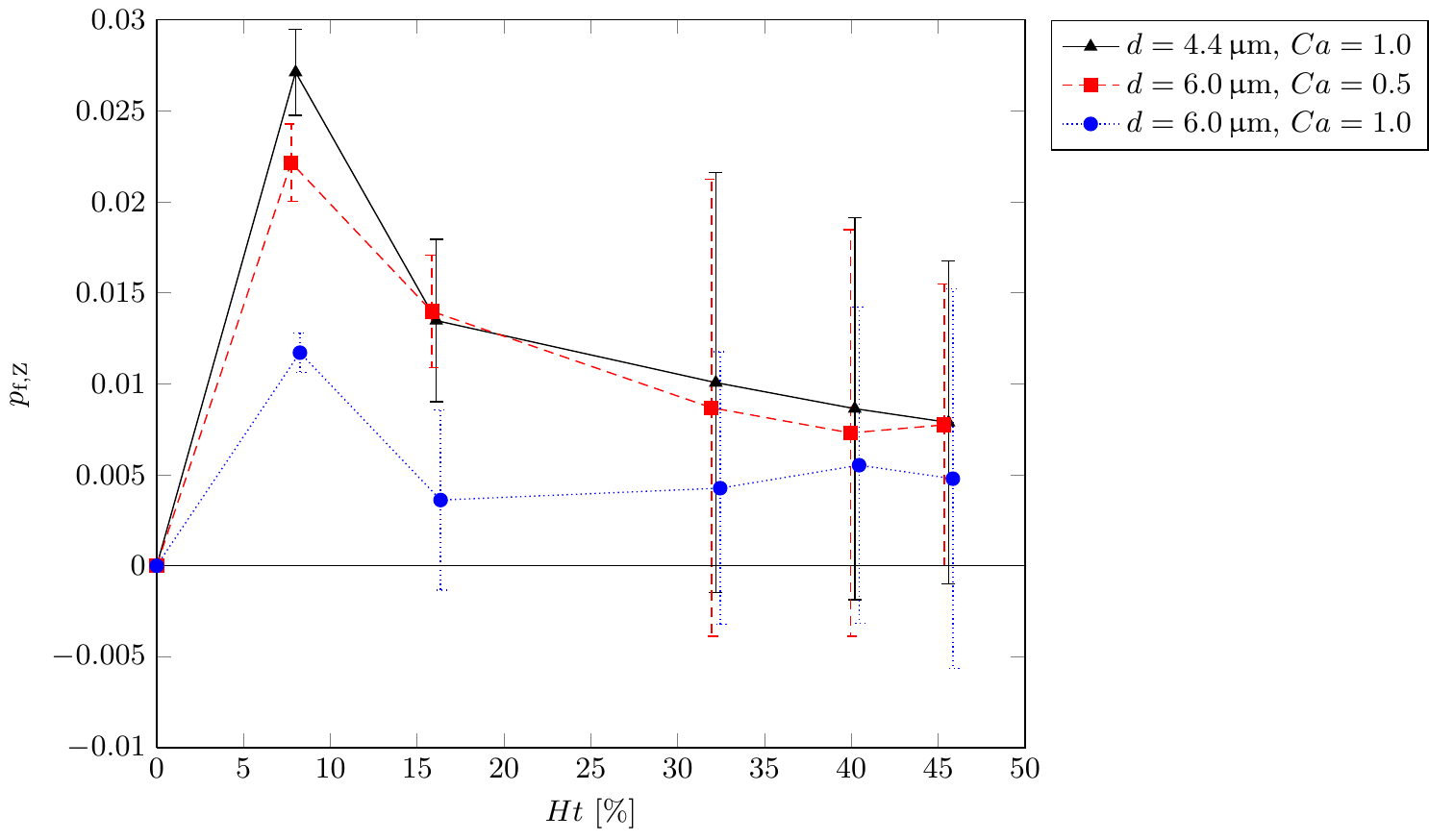}
 \caption{\label{fig:zgz_fail} Zigzag failure probability $p_\text{f,Z}$ as function of the volume fraction $Ht$ for different parameters $d$ and $Ca$. Error bars indicate the standard deviation obtained from the cell ensemble in each simulation. Data points are slightly shifted along the $Ht$-axis to avoid overlap of error bars. Lines connecting the data points are guides for the eyes.}
\end{figure}

Fig.~\ref{fig:zgz_fail} depicts the average zigzag failure probability $p_\text{f,Z}$ as function of $Ht$ for various parameters ($d$, $Ca$).
The error bars in Fig.~\ref{fig:zgz_fail} show the standard deviation as obtained from the cell ensemble and therefore represent the cell scatter at the outlet.
We see that the net failure probability $p_\text{f,Z}$ is small and remains below $3\%$ for all investigated data points.
We also observe that the failure probability increases sharply with a small increase in the volume fraction and then, with further increase in $Ht$, drops to a near constant value at high volume fractions.
This underlines the relative stability of the zigzag mode at high volume fractions.

The reason for this stability is that at higher volume fractions the RBCs tend to occupy most of the operational volume in the device.
Their behaviour is therefore akin to the fluid molecules flowing through the device, which is the natural state for a DLD device.
The relatively densely packed RBCs therefore tend to follow the natural division of streamlines.

The standard deviation of $p_\text{f,Z}$ is seen to increase with $Ht$.
This is probably due to the related increase of the collision rate between cells.
For a practical device this suggests the requirement for an outlet with greater width for RBC collection.

Cell collisions tend to make the system less deterministic.
We expect that systems with parameters $(d, Ca)$ close to the separation line in Fig.~\ref{fig_phasespace} are less robust upon an increase of $Ht$ than those systems which are farther away from the separation line.
Indeed, from Fig.~\ref{fig:zgz_fail} we see that the data point ($d = 6.0\,\micron$, $Ca = 1.0$) leads to a lower failure rate than the other two data points which are closer to the separation line in Fig.~\ref{fig_phasespace}.
This interpretation will be again corroborated in section \ref{section_displacement}.
We conclude that, at larger volume fractions and therefore lower determinacy, there is no longer a sharp separation between displacement and zigzag modes.
Therefore, the critical separation line loses its original meaning and can only be used as a reference.

\subsubsection{Displacement failure}
\label{section_displacement}

For a single cell in an ideal displacement mode, all pillar encounters lead to a bumping event and we have $N_\text{dn}^0 = 0$ and $N_\text{up}^0 = N$.
Therefore, there cannot be more bumping events as ideally expected, so $p^+_\text{f,D} = 0$.
This means that there is only one failure mode and only one corresponding failure probability $p_\text{f,D} = p^-_\text{f,D}$.
We can write the displacement error with respect to the expected lateral position as
\begin{equation}
 \Delta_y = - N \uplambda p_\text{f,D}.
\end{equation}
Note the minus sign indicating that a displacement failure always leads to a downward motion, toward the horizontal.
This can also be written in terms of events as
\begin{equation}
 p_\text{f,D} = \frac{N_\text{up}^0 - N_\text{up}}{N} = \frac{N_\text{dn} - N_\text{dn}^0}{N} = \frac{N_\text{dn}}{N}.
\end{equation}
As for the zigzag mode, we compute the displacement failure probability for individual cells, which gives us access to the mean and standard deviation of the cell ensemble.

\begin{figure}
 \centering
 \includegraphics[width=0.8\linewidth]{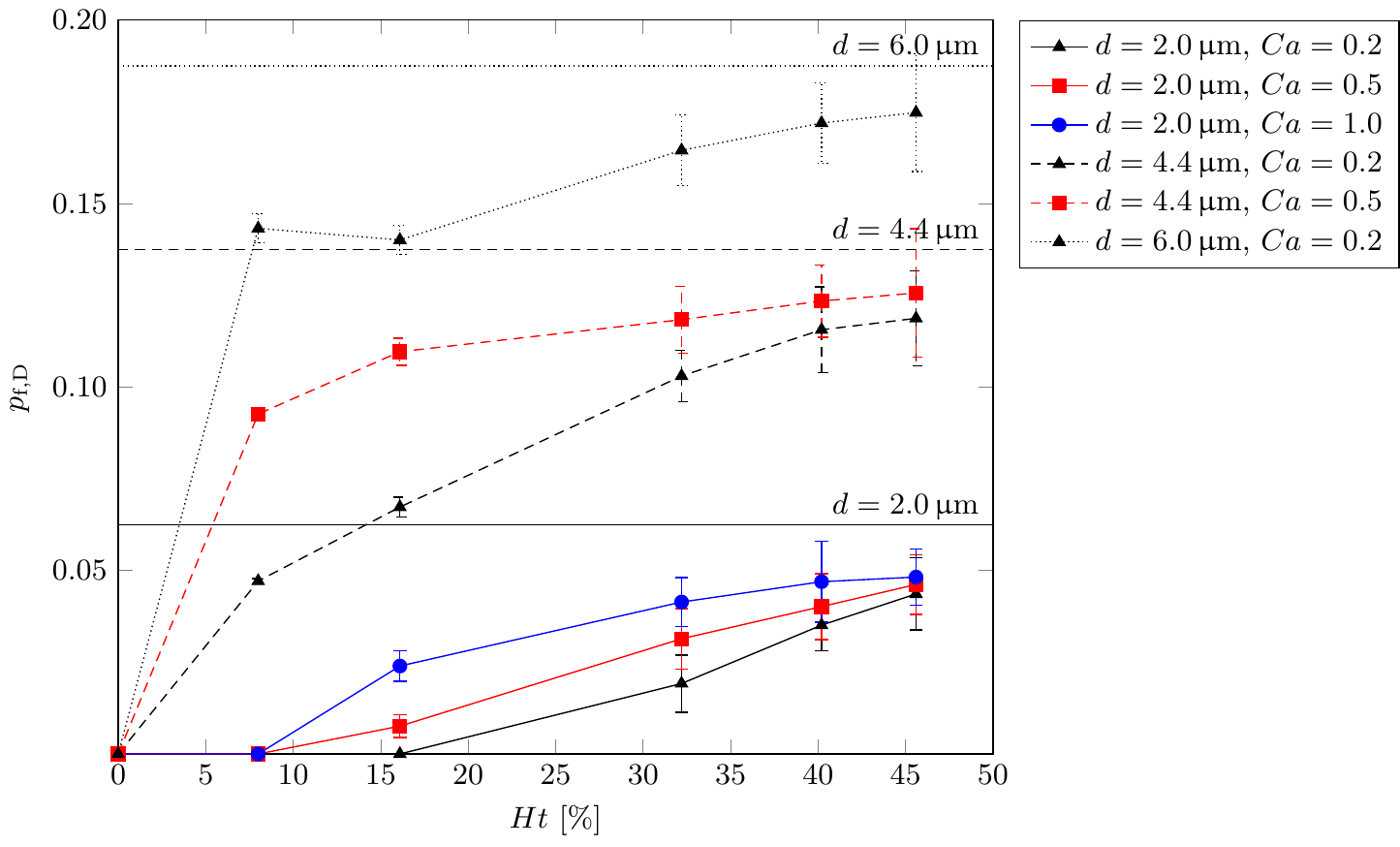}
 \caption{\label{fig:dsp_fail} Displacement failure probability $p_\text{f,D}$ as function of the volume fraction $Ht$ for different parameters $d$ and $Ca$. Error bars indicate the standard deviation obtained from the cell ensemble in each simulation. Lines connecting the data points are guides for the eyes. The horizontal lines indicate the limiting displacement failure probability $p^\infty_\text{f,D} = d / \uplambda$ for each value of $d$ (solid for $d = 2.0\,\micron$, dashed for $d = 4.4\,\micron$ and dotted for $d = 6.0\,\micron$).}
\end{figure}

In Fig.~\ref{fig:dsp_fail} the average displacement failure probabilities are plotted as function of $Ht$ for different combinations of $d$ and $Ca$.
In addition, the \emph{limiting displacement failure probability} $p^\infty_\text{f,D}$ for each row shift $d$ is shown as a horizontal line.
$p^\infty_\text{f,D}$ corresponds to the case where the cells effectively move along the horizontal and is given by $p^\infty_\text{f,D} = d / \uplambda$.
This is equivalent to the complete breakdown of the displacement mode and obviously the worst-case scenario for a DLD device intended to be run in displacement mode.

From Fig.~\ref{fig:dsp_fail} we see that $p_\text{f,D}$ increases monotonically with $Ht$ for all combinations of $d$ and $Ca$ and asymptotically approaches $p^\infty_\text{f,D}$. The highest failure rate is observed at the maximum haematocrit value studied ($Ht = 45.6\%$). In this scenario, we find RBCs nearly everywhere in the DLD device. As explained in the previous section, under these circumstances, the cellular flow closely resembles the behaviour of a continuum fluid in the device. When the cells become densely packed, not all of them can undergo displacement at every pillar crossed.
Every cell missing a displacement event increases the failure probability. Therefore, at high volume fractions, the cells tend to approach the horizontal and move on near-zigzag trajectories, which in turn leads to the breakdown of the displacement mode.

At lower volume fractions (Fig.~\ref{fig:dsp_fail}), $Ca$ has a significant effect on $p_\text{f,D}$ for fixed $d$.
In contrast, at higher volume fractions, the probability $p_\text{f,D}$ converges for various $Ca$-values at given row shifts $d$.
Therefore, at high volume fractions, the failure probability is no longer a function of $Ca$.
This shows that effects due to dense cell packing become more important than cell deformability.

Generally, the variation of $p_\text{f,D}$ with $d$ and $Ca$ can be correlated with the distance of the ($d$, $Ca$) point from the critical separation line in Fig.~\ref{fig_phasespace}.
For the zigzag case in section \ref{section_zigzag}, we found that points closer to the critical separation line have higher values of $p_\text{f,Z}$.
Here, for the displacement mode, the same argument holds, as long as $p_\text{f,D}$ is normalised by its asymptotical value $p^\infty_\text{f,D}$.
This indicates again that those systems which are closer to the critical separation line are less robust upon an increase of $Ht$.

We also see from Fig.~\ref{fig:dsp_fail} that the failure probability $p_\text{f,D}$ increases with $d$ at all volume fractions.
Therefore, we postulate that failure events in the displacement mode generally become rarer with decrease in row shift $d$.
Why is this the case?
For larger values of $d$, the critical particle separation radius becomes larger (\emph{cf.}~Tab.~\ref{tab:streamlines}).
Therefore, there are more particles in the device that are potentially close to the streamlines passing below the next pillar.
This means that more particles can accidentally be pushed by neighbouring particles onto such a streamline and subsequently fall down to a lower lane.

It is misleading to claim that a smaller value of $d$ will make DLD devices more robust, though.
DLD devices with small $d$ are typically longer, \ie~$N$ is larger.
This is so because at small $d$ a particle has to be displaced many times in order to achieve a significant lateral displacement at the outlet.
We can assume that the required $N$ for separation typically scales with $1/d$.
It is important to note that the relevant failure indicator is the product $p_\text{f,D} N$.
If $N$ grows faster than $p_\text{f,D}$ decreases upon decreasing $d$, the device will not be more robust in the end.
Therefore, a design criterion for an improved DLD device operating at large $Ht$ is to reduce the product $p_\text{f,D} N$.

\subsection{Outlet distributions}
\label{section_outlets}

\begin{figure}
 \includegraphics[width=\linewidth]{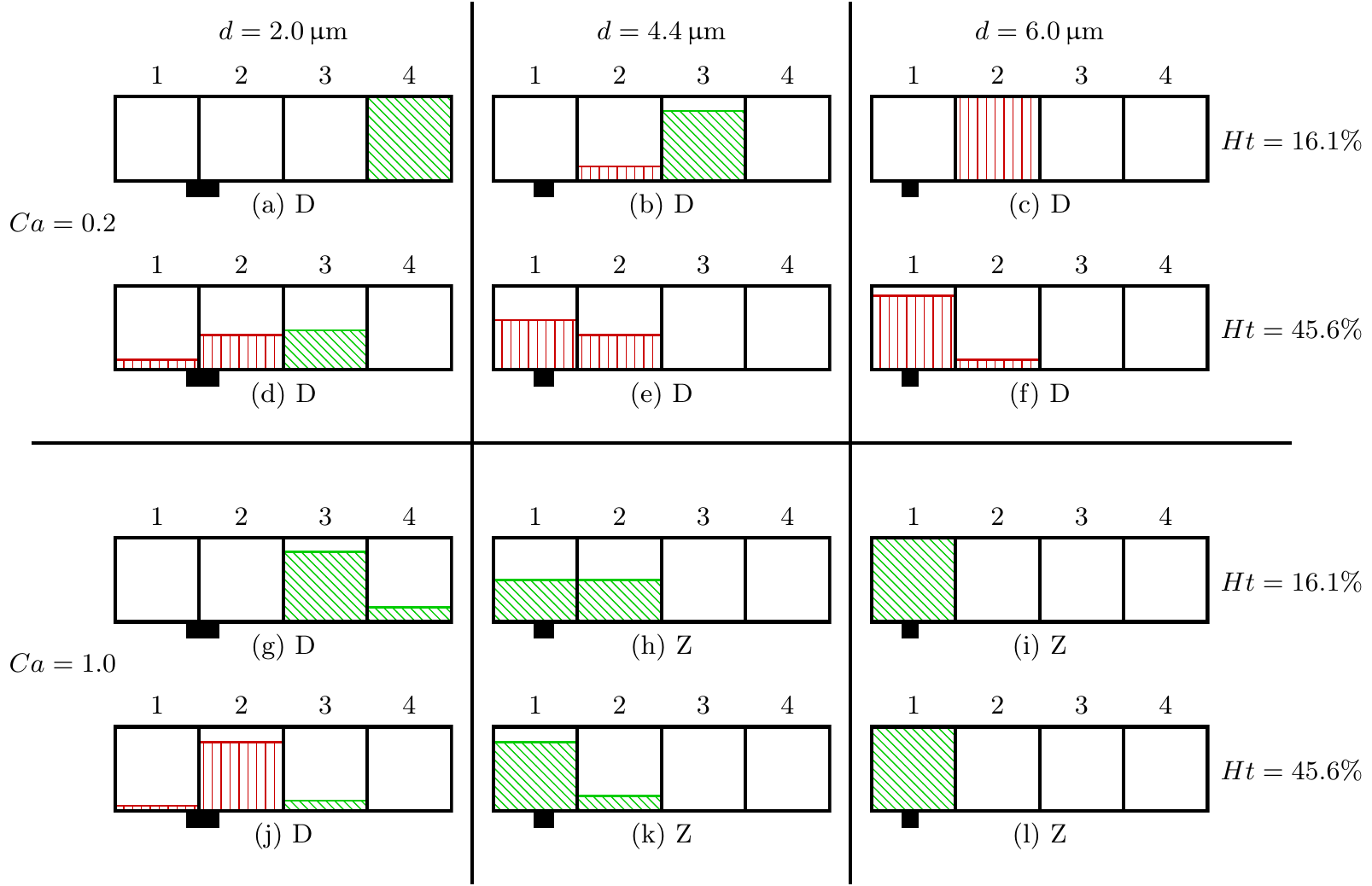}
 \caption{\label{fig_histograms} Histograms of RBC outlet distributions for various control parameters. The row shift assumes the values $d = 2.0\,\micron$ (left column), $d = 4.4\,\micron$ (middle column) and $d = 6.0\,\micron$ (right column).
 The first and second rows show the results for $Ca = 0.2$, the third and fourth rows the results for $Ca = 1.0$. The volume fraction is $Ht = 16.1\%$ (first and third rows) and $Ht = 45.6\%$ (second and fourth rows). The expected operation mode is indicated (D for displacement or Z for zigzag, \emph{cf.}~Fig.~\ref{fig_phasespace}). The inlet position relative to the total outlet width is marked as a black box at the bottom of each histogram (\emph{cf.}~Fig.~\ref{fig_outlets}). Outlets 1 and 2 are supposed to collect zigzag cells, outlets 3 and 4 displaced cells. Slanted hatching with green colour indicates successful and red-coloured vertical hatching erroneous collection of cells into their expected outlets.}
\end{figure}

The failure probabilities by themselves do not present a complete picture of the effectiveness of the device.
To reveal the robustness of the device we count the number of cells in each of the four outlets (\emph{cf.}~section \ref{subsection_simulations}) and show the cell count histograms in Fig.~\ref{fig_histograms}.
We present the histograms for $Ht = 16.1\%$ and $45.6\%$ at $d = 0.2$, $4.4$, $6.0\,\micron$ and $Ca = 0.2$, $1.0$.
For the sake of a simpler comparison, the histograms are normalised by the maximum outlet width and the total number of cells collected at the outlets.

We consider the DLD device as robust if all cells supposed to travel on zigzag trajectories are collected in outlets 1 and 2, whereas all cells expected to be displaced are collected in outlets 3 and 4.
This definition would still allow for successful separation of particles in a real-world device.
Note that the values of $d$ and $Ca$ already decide whether the cells are supposed to be displaced or not (\emph{cf.}~Fig.~\ref{fig_phasespace}).

We observe that, for the zigzag mode at $Ht = 16.1\%$ (Fig.~\ref{fig_histograms}(h,i)), all RBCs are collected in their designated outlets 1 and 2.
When we increase the volume fraction to $Ht = 45.6\%$ (Fig.~\ref{fig_histograms}(k,l)), still all cells end up in the expected outlets 1 and 2.
This may allow for larger particles to be displaced into outlets 3 and 4 from a background of RBCs even at high volume fractions.

However, the RBCs in the displacement mode reveal a different picture.
In Fig.~\ref{fig_histograms}(a,b,g), at $Ht = 16.1\%$, (nearly) all RBCs are displaced into their designated outlets.
But, when the row shift $d$ is increased to $6.0\,\micron$, there is total failure to obtain RBCs in outlets 3 and 4 (Fig.~\ref{fig_histograms}(c)) even at rather dilute $16.1\%$ volume fraction.
With an increase in the volume fraction to $45.6\%$, we get about $50\%$ correctly collected RBCs at a low row shift, $d = 2.0 \micron$.
This deteriorates quickly when $d$ is increased, with all cells collected in the zigzag outlets (1 and 2) at $d = 4.4$ and $6.0\,\micron$ (Fig.~\ref{fig_histograms}(e,f)).

This evidence corroborates our postulate that a DLD device lends relatively easily to separation of larger particles (such as white blood cells) from a background of RBCs designated to be in the zigzag mode.
However, it will be more difficult to achieve efficient separation at high volume fractions when the RBCs are expected to be the displaced particle species in order to separate smaller particles (such as platelets).

Fig.~\ref{fig:dsp_fail} and Fig.~\ref{fig_histograms} suggest that already small volume fractions, around $10\%$, can significantly spoil the displacement mode efficiency, at least for larger values of $d$.
The large displacement failure probability $p_\text{f,D}$ is the reason for this behaviour.
Since $p_\text{f,D}$ is governed by cell collisions which reduce the determinacy of the device, we believe that this problem is not specific to the considered device geometry and particle type (RBCs).
It is rather a general problem for all devices which rely on some kind of deterministic displacement of any type of particles (not only RBCs) interacting with obstacles.
We therefore expect that most DLD devices will not work reliably at large particle volume fractions.

Inglis et al.~\cite{inglis_scaling_2011} experimentally observed a decreasing white blood cell enrichment efficiency with increasing haematocrit in a DLD device (see Fig.~6 in \cite{inglis_scaling_2011}).
Apart from this result, we did not find discussions of the haematocrit effect on the separation efficiency of blood components in the literature.

It is an open question how to keep the advantages of a DLD device (\emph{deterministic} displacement of large particles) and at the same time avoid displacement failures caused by \emph{non-deterministic} particle collisions.
Therefore, a key requirement for more robust DLD devices is to reduce the displacement failure probability $p_\text{f,D}$ at larger volume fractions.
This may be achieved by optimising the pillar shape or the size of the gaps between the pillars.
We believe that DLD devices will not be suitable for suspensions with high volume fractions (above about $10$--$20\%$) before this problem is solved.

\section{Summary and conclusions}
\label{section_conclusions}

Deterministic lateral displacement (DLD) devices are commonly used in microfluidics to separate particles based on their size or deformability, \eg~cells in whole blood.
Depending on its size, a particle experiences one of two possible trajectory modes in a DLD device: ``zigzag'', where a small particle follows the fluid streamlines on average, and ``displacement'', where larger particles bump into obstacles in the flow and are forced on displaced trajectories.
However, DLD devices are designed for the treatment of relatively dilute suspensions.

In the present work, we examine the effect of red blood cell (RBC) volume fraction (haematocrit $Ht$) on the performance of a DLD device \emph{via}
3D simulations based on the immersed-boundary, lattice-Boltzmann and finite-element methods.
In order to quantify failure in each mode, we analyse displacement and zigzag failure probabilities.
A failure event denotes a cell encounter with a DLD obstacle which leads to a ``wrong'' outcome, \ie~the cell moves in a different lateral direction than expected for a dilute suspension in the same device geometry.
We find that the mean and standard deviations of these failure probabilities are significant performance indicators for a DLD device.

Our main observation is that the displacement mode breaks down upon an increase of the RBC volume fraction, caused by large failure probabilities.
At the same time, the zigzag mode shows relatively few failure events and is more robust at higher volume fraction.
We find that, in contrast to the displacement failure, the mean zigzag failure probability seems largely independent of the haematocrit.
This difference stems from the fact that, in the zigzag mode, a cell suffers from two failure modes which tend to cancel each other.
In the displacement mode, however, there exists only one failure mode whose effect accumulates over time.

Furthermore, we investigate at which lateral position the simulated RBCs are found when they reach the end of the device.
We define four outlet bins and analyse the cell count in each outlet.
We observe that RBCs expected in the zigzag mode are essentially collected in the designated zigzag outlets, while RBCs meant to travel on displacement trajectories fail to be collected in displacement outlets at volume fractions above about 10--20\%.
As a consequence, it seems to be easier to separate larger particles (\eg~white blood cells) from a dense RBC background than smaller particles (\eg~platelets) from RBCs.

The essential reason for the breakdown of the displacement mode at larger $Ht$ is that DLD devices rely on \emph{deterministic} processes while particle collisions in dense suspensions are intrinsically \emph{non-deterministic}.
We therefore believe that our results are not a peculiarity of the specific DLD geometry and choice of red blood cells studied here, but that basically all separation devices relying on deterministic displacement would suffer from any dense suspension effects.
This would make it difficult, if not impossible, to separate dense suspensions in DLD devices without prior diluting.

The key for future applications of the DLD technique for dense suspensions, blood in particular, is to understand how the failure probabilities are affected by particle collisions and how this effect may be reduced, \eg~by novel obstacle shapes.

\section*{Acknowledgement}
  
T.K.~thanks the University of Edinburgh for the award of a Chancellor's Fellowship and full funding of this work.
We thank David Inglis for stimulating discussions.
There are no conflicts of interest.
Ethical approval was not required.

\section*{Bibliography}

\bibliographystyle{elsarticle-num}
\bibliography{bibliography}
  
\end{document}